\documentclass[preprint2]{aastex}
\usepackage{graphicx}
\usepackage{float}

\slugcomment{Draft on \today}

\shorttitle{J1211+1437 a hyper-velocity subdwarf binary}
\shortauthors{N\'{e}meth et al.}

\begin{document}


\title{An extremely fast halo hot subdwarf star in a wide binary system}


\author{
P\'{e}ter~N\'{e}meth\altaffilmark{1},
Eva~Ziegerer\altaffilmark{1},
Andreas~Irrgang\altaffilmark{1},
Stephan~Geier\altaffilmark{1,2},
Felix~F\"urst\altaffilmark{3},
Thomas~Kupfer\altaffilmark{3,4},
Ulrich~Heber\altaffilmark{1}
}

\affil{
$^1$Dr.~Karl~Remeis-Observatory \& ECAP, Astronomical Institute,
Friedrich-Alexander University Erlangen-Nuremberg,
Sternwartstr.~7, D-96049 Bamberg, Germany\\
$^2$Department of Physics, University of Warwick, Coventry, CV4 7AL, UK\\
$^3$Division of Physics, Mathematics, and Astronomy, California Institute
of Technology, Pasadena, CA 91125, USA\\
$^4$Department of Astrophysics/IMAPP, Radboud University Nijmegen, P.O. Box
9010, 6500 GL Nijmegen, The Netherlands\\
}




\begin{abstract}
New spectroscopic observations of the halo hyper-velocity 
star candidate
SDSS\,J121150.27+143716.2 ($V=17.92$ mag) revealed a cool companion 
to the hot subdwarf primary. 
The components have a very similar radial velocity and their 
absolute luminosities
are consistent with the same distance, confirming the physical nature of the 
binary, which is the first double-lined hyper-velocity 
candidate.
Our spectral decomposition of the Keck/ESI spectrum
provided an sdB+K3V pair, analogous to
many long-period subdwarf binaries observed in the Galactic disk. 
We found the subdwarf atmospheric parameters: 
$T_{\rm eff}=30\,600\pm500$\,K,
$\log{g}=5.57\pm0.06$\,cm\,s$^{-2}$ and He abundance 
$\log(n{\rm He}/n{\rm H})=-3.0\pm0.2$. 
Oxygen is the most abundant metal in
the hot subdwarf atmosphere, and Mg and Na lines are the most 
prominent spectral features of the cool companion,
consistent with a metallicity of $[{\rm Fe}/{\rm H}]=-1.3$.
The non-detection of radial velocity variations suggest the orbital period 
to be a few hundred days, in agreement with similar binaries observed in the disk. 
Using the SDSS-III flux calibrated spectrum we measured the distance 
to the system $d=5.5\pm0.5$\,kpc, which is
consistent with ultraviolet, optical, and infrared photometric constraints 
derived from binary spectral energy distributions.  
Our kinematic study shows that the Galactic rest-frame velocity of the
system is so high that an unbound orbit cannot be ruled out. 
On the other hand, a bound orbit requires a massive dark matter halo. 
We conclude that the binary either formed in the halo
or it was accreted from the tidal debris of a dwarf galaxy by the Milky Way.
\end{abstract}





\section{Introduction \label{sec:intro}}

The stellar population in the Galactic halo is distinct from the populations
in the thin and thick disks. 
Without evidence for ongoing star formation, halo stars are in general
metal-poor and hence much older than their disk counterparts. 
Those so-called population II stars also show peculiar kinematics.
While disk stars are orbiting around the Galactic Center with 
velocities around $240\,{\rm km\,s^{-1}}$ for the solar neighborhood,
the orbits of halo stars are more diverse, usually tilted against 
the disk and sometimes retrograde. 
The typical distribution for their three-dimensional
Galactic rest-frame velocities
($v_{\rm grf}$) ranges from about $-200\,{\rm km\,s^{-1}}$ to 
$+200\,{\rm km\,s^{-1}}$ (e.g. \citealt{brown07}).

Objects with even higher $v_{\rm grf}$
and peculiar kinematics are known among halo stars.
Runaway stars reach velocities of up to $\sim300\,{\rm km\,s^{-1}}$, while
the so-called hyper-velocity stars (HVS) travel with velocities, that can
even exceed the escape velocity of the Galaxy (\citealt{brown05}; 
\citealt{hirsch05}; \citealt{edelmann05}) and reach
more than $\sim1000\,{\rm km\,s^{-1}}$ \citep{geier15}. 

Since most of the known runaway and HVS stars are young, early-type 
main-sequence stars, they cannot belong
to the old halo population (see \citealt{brown15} for a review). 
Those objects must have been formed in the Galactic disk, accelerated, and
eventually injected into the halo. 
However, the acceleration mechanisms still remain unclear.
Runaway stars might be the remnants of binary systems where the more
massive
component exploded as a core-collapse supernova (SN, \citealt{blaauw61})
or they might have been ejected by dynamical interactions in dense star
clusters.
HVS stars can be generated when a close binary is disrupted by
the supermassive black hole (SMBH) in the center of our Galaxy 
and one component is ejected \citep{hills88}.

A peculiar sub-population of these extreme halo stars consists of 
O-type and B-type hot subdwarf stars (sdO/Bs; see \citealt{heber09} for a review). 
One of those stars (US\,708, HVS\,2) turned out to be the fastest unbound
star in our Galaxy (\citealt{hirsch05}; \citealt{geier15}).   
Hot subdwarfs form if the progenitor loses its envelope almost    
entirely after passing the red giant branch and the remaining hydrogen-rich   
envelope has not retained enough mass to sustain a hydrogen-burning shell.    
Mass-transfer in a binary system is considered to be the most likely
mechanism
to remove the envelope and form a hot subdwarf \citep{han02,han03}.
It is challenging to explain the acceleration of such stars to high
velocities,
because the most plausible scenarios are only 
able to explain the acceleration of single stars. 

\cite{geier15} explained the extreme velocity of US\,708 by
proposing that this star is the ejected companion of a very tight binary
system where the massive white dwarf primary exploded as a 
thermonuclear type Ia SN.
In this scenario mass-transfer from the hot subdwarf to the white dwarf
triggered the SN explosion. 
Since the properties of such ejected companions allow us to put constraints
on the yet unknown parameters of the progenitor systems of SNe Ia, it is    
crucial to find more objects like US\,708.

Several other hot subdwarfs with high $v_{\rm grf}$ have
been discovered by \cite{tillich11}.  
The fastest object in this sample is the sdB star
SDSS\,J121150.27+143716.2 (also PB\,3877; here J1211 for short) 
with $v_{\rm grf}=713\pm140\,{\rm\,km\,s^{-1}}$, which is predominantly a
tangential velocity.
\cite{tillich11} found J1211 to most likely be unbound to the Galaxy.
They traced the trajectory of this star back to the outskirts 
of the Galactic disk and showed that this star could not have been ejected 
by the SMBH in the Galactic Center.
Those properties made J1211 an excellent candidate for the SN Ia
ejection scenario and a prime target for our follow-up campaign with 
8m-class telescopes.

\section{Observations and spectroscopic analysis \label{sec:observations}}


\begin{figure*}
\includegraphics[width=\hsize]{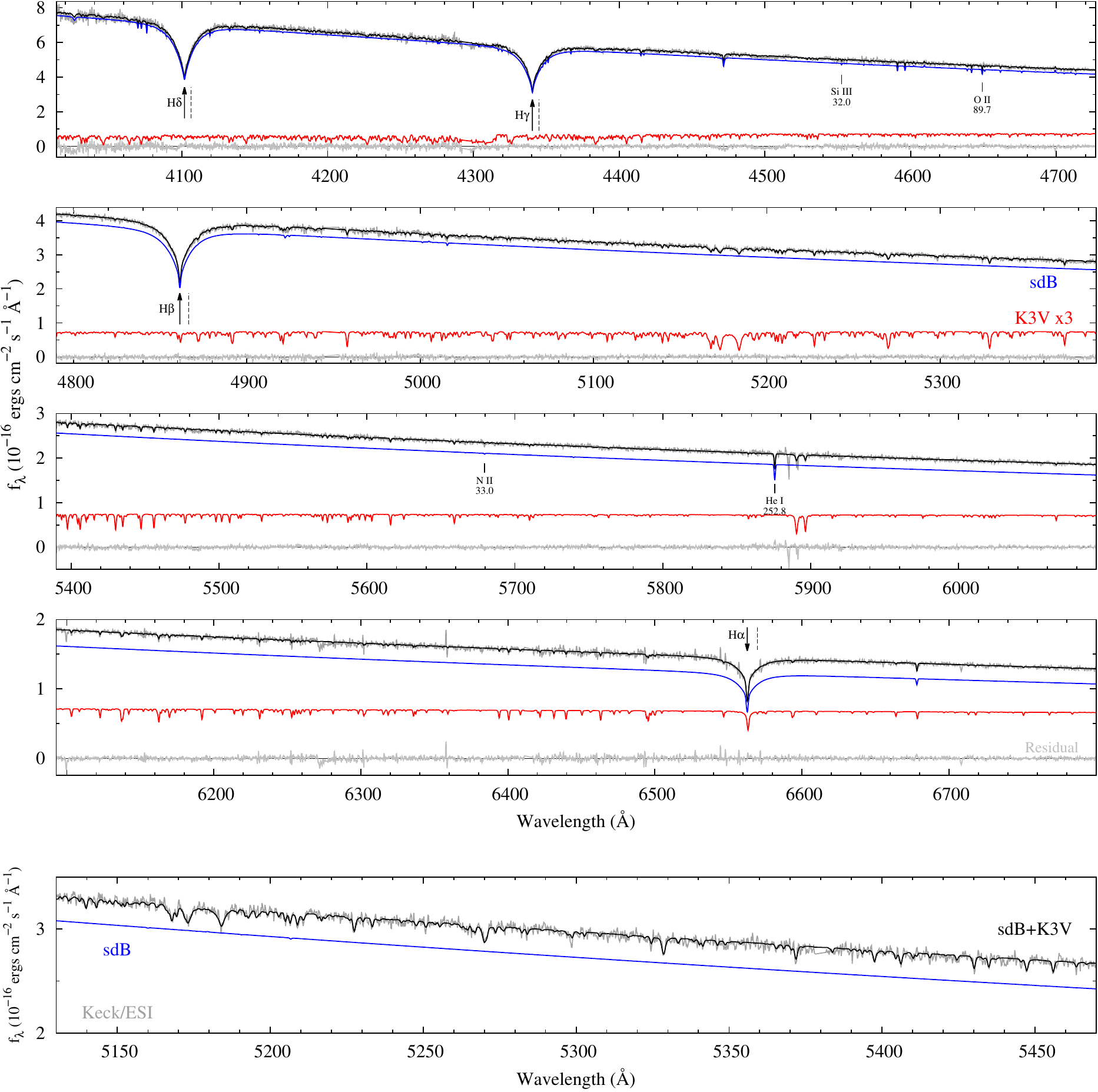}
\caption{
{\sl Top panels:} 
Radial-velocity-corrected and coadded Keck/ESI spectrum of J1211
(gray).
The subdwarf (blue) dominates the spectrum in
the entire range, while the K3V companion (red) contributes 
$17\pm3$ \% of the flux at 6750 \AA. 
The thin black line overplotted in the observation is the
best-fit {\sc XTgrid} binary model, the sum of the two
components.
To make the companion distinguishable from the residual its flux is
multiplied by three. 
The flux level was adjusted to the {\sc BOSS/DR12} spectrum.
The dashed lines indicate the observed wavelengths of the Balmer lines. 
The same shift can be observed between the Na D lines originating 
from the companion and from the interstellar material. 
The strongest He, N, O and Si lines are marked and labeled with their
theoretical equivalent widths in m\AA. 
The atmospheric parameters of the binary members are given in
Table\,\ref{tbl:one}. 
{\sl Bottom panel:} zoom-in to the strongest lines of the companion 
in the $5150-5450$ \AA\ region. 
}
\label{fig:1} 
\end{figure*}

To improve the atmospheric parameters of J1211 and measure the
rotational 
broadening of the lines, we obtained medium-resolution spectra with the ESI 
spectrograph ($R=8000$, $4000-6800\,{\rm \AA}$) at the Keck telescope and
the 
XSHOOTER spectrograph ($R=10\,000$, $3000-10\,000\,{\rm \AA}$) at the ESO-VLT 
between 2014 June 1 and 11. 
To our surprise, the better resolutions and S/Ns of these spectra, compared to 
the SDSS and ESO-VLT FORS1 spectra used by \cite{tillich11}, 
immediately revealed the unseen weak lines of a cool companion in the
spectrum.

We performed a model atmosphere analysis and fitted the composite spectra
with the steepest-descent spectral analysis program {\sc XTgrid} 
\citep{nemeth12}. 
This fit procedure allows the separation of the members
in composite spectra binaries and 
employs {\sc Tlusty/Synspec} (\citealt{hubeny95}; 
\citealt{lanz07}) model atmospheres for the hot subdwarf, and 
{\sc Atlas9} models for the cool companion. 
The plane-parallel, non-LTE {\sc Tlusty} model atmospheres are calculated 
in a hydrostatic and radiative equilibrium that is appropriate for hot
subdwarfs. 
Our models include H, He, C, N, O and Si opacities that are 
consistently in the model 
atmosphere calculations and in the spectral synthesis. 
{\sc XTgrid} starts with an initial model and iteratively updates the 
parameters of the binary along the steepest-gradient of the global
chi-squared ($\chi^2$) surface. 
{\sc Tlusty} models are calculated in every iteration and the spectrum of the 
cool companion is regularly extracted from the {\sc BlueRed} high-resolution
synthetic stellar spectral library \citep{bertone08}. 
The {\sc BlueRed} grid allows to fit the temperature, gravity, and
scaled-solar metallicity of the cool companion using interpolated models. 
The fit was based only on spectral lines because a reliable flux calibration 
was not available for the higher quality Keck and VLT data. 
The best fit is shown in Fig.~\ref{fig:1} and the final parameters
together with error bars are listed in Table\,\ref{tbl:one}. 
For a better representation of the observed data the continuum of the Keck 
observation  was adjusted to the composite model in Fig.\ref{fig:1}.
No attempt has been made to include parameter correlations in the error
calculations.

Our fit confirmed the sdB classification of the primary and provided a K3V
type main-sequence companion.
The revised sdB temperature is systematically lower by $\sim$2000 K, while 
the surface gravity and He abundance are within error bars of the 
previous analysis by \cite{tillich11}. 
The discrepancy originates from the different data set used in the analysis. 
\cite{tillich11} used the SDSS-I DR7 spectrum that shows significant 
differences compared to the SDSS-III, Keck, and VLT data analyzed here.
We note that the $\chi^2$ map of a composite spectrum model
resembles a valley, or canyon, with a flat bottom due to 
strong correlations between the atmospheric and binary parameters.
Therefore, a 
$T_{\rm eff}=28\,000$ K, $\log{g}=5.46$ cm\,s$^{-2}$ or a
$T_{\rm eff}=33\,000$ K, $\log{g}=5.67$ cm\,s$^{-2}$ solution 
gives only a 10\% higher $\chi^2$ minimum. 
Similar degeneracies have been reported in other composite spectra
binaries \citep{vos13}.

It is not possible to determine the orbital parameters of the binary system
with the available data. 
We measured the radial velocities (RVs) of two SDSS spectra taken in 2005
and 2012, an ESO-VLT FORS1 spectrum taken in 2008, seven ESI spectra, and 
eight XSHOOTER spectra taken in 2014. 
No significant RV shift was detected over the timebase of nine years. 
Furthermore, the RV from the spectral lines of the sdB primary closely
matches the one of the K-star; they both are shifted 
by $\sim240$\,km\,s$^{-1}$ with respect to the barycentric reference frame. 
A chance alignment can therefore be excluded and the two components visible
in the spectra indeed form a binary system.

About 20-30\% of sdB stars are in composite spectrum binaries with F-G-K
type companions. 
The orbital periods for about a dozen of these systems could 
be determined only recently (e.g. \citealt{vos13}).
They turned out to lie between $\sim$700 and $\sim$1300 days, which is
consistent with the Roche-lobe overflow evolution channel \citep{han02}.
Stable mass transfer to the main-sequence companion most likely removed the 
envelope of the red giant and led to the formation of the sdB \citep{chen13}. 
Typical RV semi-amplitudes of those binaries are smaller than 
$20\,{\rm km\,s^{-1}}$. 
Although the low resolution of our data did not allow us to find RV variations, 
the RVs of the sdB and the K-star differ by 
$25\pm13$\,km\,s$^{-1}$ in the coadded Keck spectrum,
which is consistent with such wide binaries.
Therefore, assuming an inclination of $i>30^\circ$, we expect
that J1211 is a typical sdB+MS binary with an 
orbital period of several hundred days and a separation of several AU.
An observational confirmation will require high dispersion spectroscopy.

\begin{table*} \small
\caption[]{Parameters of the Spectroscopic Model Fit Shown in 
Figure\,\ref{fig:1}, 
Photometric Model Shown in Figure\,\ref{fig:2}, and Kinematic Model
Shown in Figure\,\ref{fig:3}.
} 
\label{tbl:one}
\centering
\renewcommand{\arraystretch}{1.1}
\begin{tabular}{l|p{1.8cm}|p{1.8cm}|p{1.8cm}|p{1.8cm}}
\hline\hline 
%
{\bf Spectroscopic Parameters}            & {\bf Value}  & \boldmath{$+1 \sigma$} & \boldmath{$-1 \sigma$} &\boldmath{$\times$}\,{\bf Solar}\\
\hline
{\sl sdB:}&&&&\\
$T_{\rm eff}$ (K)                   & 30600   & 400 & 500 &       \\
$\log{g}$ (cm\,s$^{-2}$)            & 5.57    & 0.08& 0.06&       \\
$\log n(\mathrm{He})/n(\mathrm{H})$ (dex) &--3.02   & 0.20& 0.10&0.009 \\
$\log n(\mathrm{C})/n(\mathrm{H})$ (dex)  &--$4.78>$&     &     &0.037$>$\\
$\log n(\mathrm{N})/n(\mathrm{H})$ (dex)  &--4.60   & 0.16& 0.17&0.370 \\
$\log n(\mathrm{O})/n(\mathrm{H})$ (dex)  &--4.47   & 0.14& 0.25&0.069 \\
$\log n(\mathrm{Si})/n(\mathrm{H})$ (dex) &--5.80   & 0.19& 1.15&0.049 \\
\rule{0pt}{4ex}{\sl K3V:}&&&&\\
$T_{\rm eff}$ (K)                   & 4850    & 300 & 300 & \\
$\log{g}$ (dex)                     & 4.6     & 0.4 & 0.4 & \\
$[M/H]$   (dex)                     &--1.3    & 0.3 & 0.3 & \\
\rule{0pt}{4ex}F$_2$/F$_1$ (at 6750 \AA) &$0.189$& 0.039    & 0.027& \\
Distance (kpc)                      & 5.4    & 0.5 & 0.5 & \\
\hline\hline
{\bf Photometric Parameters}    & {\bf Value}  & \boldmath{$+1 \sigma$} & \boldmath{$-1 \sigma$} & \\
\hline
{\sl K3V:}&&&&\\
$T_{\rm eff}$ (K)                   & 4610    & 180  & 200  & \\ 
$\log{g}$ (dex)                     & 4.51    & 0.14 & 0.04 & \\ 
$M$   (M$_\odot$)         & 0.7     & 0.2 & 0.0 & \\ 
$A_V$ (mag)                         & 0.122   & 0.015 & 0.016 & \\ 
Distance (kpc)                      & 5.68    & 0.10 & 0.09 & \\
\hline\hline
{\bf Kinematic Parameters (Model-III)} & \multicolumn{4}{c}{{\bf Values} \boldmath{$\pm$ 1$\sigma$}}\\
\hline
{\sl At the current location:}      & \multicolumn{4}{c}{} \\
Heliocentric radial and tangential velocity (km\,s$^{-1}$): & \multicolumn{4}{c}{ $v_r=234.5\pm2.1$; $v_t=520\pm$86} \\
Proper motion (mas\,yr$^{-1}$; \citealt{tillich11}):  & \multicolumn{4}{c}{ $\mu_\alpha(\cos{\delta})=-12.1\pm$1.8; $\mu_\delta=-27.2\pm1.4$  } \\
Cartesian coordinates (kpc)   & \multicolumn{4}{c}{              $x=-8.54\pm0.02$;    $y=-1.48\pm0.14$; $z=5.35\pm0.49$} \\
Cartesian velocities (km\,s$^{-1}$)& \multicolumn{4}{c}{         $v_x=81.3\pm45.8$; $v_y=-562.4\pm77.8$; $v_z=26.6\pm21.7$} \\
Galactic radial velocity (km\,s$^{-1}$)& \multicolumn{4}{c}{     $U=16.9\pm48.3$} \\
Galactic rotational velocity (km\,s$^{-1}$)& \multicolumn{4}{c}{ $V=-567.9\pm77.1$} \\
Galactic rest-frame velocity (km\,s$^{-1}$)& \multicolumn{4}{c}{ $v_{\rm grf}=571.3\pm76.4$} \\
\rule{0pt}{4ex}{\sl At the last disk passage:}           & \multicolumn{4}{c}{}\\
Time of disk passage   (Myr)  & \multicolumn{4}{c}{              $T=-69.5\pm21.1$} \\
Cartesian coordinates  (kpc)  & \multicolumn{4}{c}{              $x=-8.3\pm3.3$; $y=34.8\pm15.1$; $z=0.0$} \\
Cartesian velocities  (km\,s$^{-1}$)  & \multicolumn{4}{c}{      $v_x=-45.2\pm40.1$; $v_y=-420.3\pm48.7$; $v_z=96.2\pm18.4$} \\
Galactic radial velocity (km\,s$^{-1}$) & \multicolumn{4}{c}{    $U=-393.2\pm66.1$} \\
Galactic rotational velocity (km\,s$^{-1}$) & \multicolumn{4}{c}{$V=-149.8\pm35.7$} \\
Galactic rest-frame velocity (km\,s$^{-1}$) & \multicolumn{4}{c}{$v_{\rm grf}=436.5\pm41.6$} \\
\hline\hline
\end{tabular}
\parbox{\textwidth}{
{\bf Note.} The quoted error bars are one dimensional statistical errors. 
Abundances are also listed with respect to the solar mixture \citep{asplund09}.
Note: The cylindrical galactic velocity component $W$ is 
equal to the Cartesian velocity component $v_z$.}
\end{table*}




\section{Spectroscopic distance\label{sec:distance}}

The neglect of the companion flux and the different flux calibration of the
SDSS-I DR7 spectrum provided a large distance in the previous analysis. 
Therefore, we calculated the distance to J1211 by scaling the 
synthetic composite spectrum to the flux calibrated {\sc SDSS-III/BOSS}
DR12 observation. 
The spectral decomposition allowed us to measure the flux ratio ($F_r$) 
from the composite spectrum:
\[
F_r=\frac{F_2}{F_1}=\frac{f_2{R_2}^2}{f_1{R_1}^2}
\]
where all flux values are monochromatic and $f_1$ is the flux of 
the subdwarf with radius $R_1$, and $f_2$ and $R_2$ 
are the surface flux and radius of the companion. 
We approximated the subdwarf mass with the canonical mass of 
0.48\,$M_\odot$ and used the surface gravity from spectroscopy 
to find the radius $R_1= 0.188\pm0.013\,R_\odot$, which is a typical 
value for similar hot subdwarfs. 
Next, we used the surface flux of the subdwarf model to find the 
absolute contribution of the companion ($F_2$). 
Then, from the ratio of the observed flux ($F$) and the composite model,
the flux scale factor ($F_1\,(1 + F_r)\,F^{-1}$) and the distance can 
be calculated:
\[
d = R_1\sqrt{ \frac{F_1(1 + F_r)}{F} }
\]
  
The different radii of the subdwarf and the cool companion are implicitly
included in the composite model and set up by the fit procedure through the
flux ratio. 
The synthetic composite spectrum based on the Keck observation 
fits the slope of the BOSS spectrum well, also suggesting a relatively 
low reddening toward J1211 in agreement with the low interstellar 
extinction in its direction (see Sect.~\ref{sect:phot}).
We measured the scale factor around 5800 \AA\ where the effects of
reddening, atmospheric extinction and spectral lines are the lowest in 
the observation. 
We found a scale factor of $1.66\pm0.05\times10^{24}$, which gives 
a spectroscopic distance of $d=5.4\pm0.5$\,kpc.

\section{Photometric consistency check}\label{sect:phot}

\begin{figure*}
\centering
\includegraphics[width=0.68\textwidth]{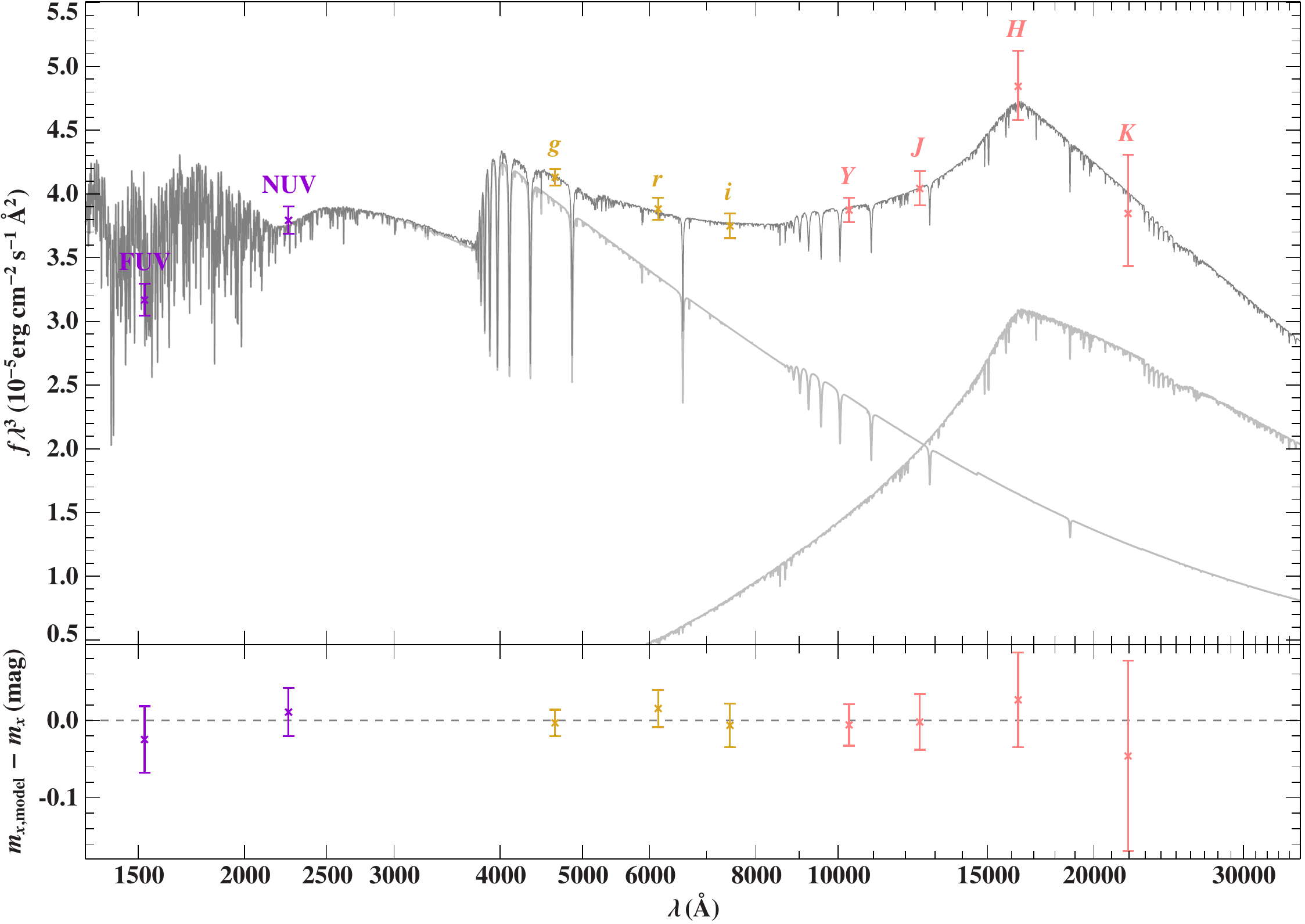}
\caption{Comparison of synthetic and observed photometry.
{\sl Top panel:} the spectral energy distribution. 
The colored data points are fluxes, which are converted from 
observed magnitudes, and the solid gray line is the composite 
(sdB+K) model. 
The individual contributions are plotted in light gray. 
{\sl Bottom panel:} the residuals show the differences 
between synthetic and observed magnitudes. 
The photometric systems have the following color code: 
GALEX (violet), SDSS (gold), UKIDSS (pink).}
\label{fig:2}
\end{figure*}

To check whether the spectroscopic results are consistent with photometry, 
we performed a fit of the observed spectral energy distribution (SED). 
The latter is given by the following magnitudes compiled from the literature: 
GALEX DR5 $\textit{FUV} = 16.663$\,mag and $\textit{NUV} = 17.039$\,mag 
\citep{bianchi11}, SDSS\footnote{The $u$- and $z$-magnitudes 
of SDSS are obvious outliers and are thus not used.} 
DR12 $g = 17.740$\,mag, $r = 18.076$\,mag, and $i =18.298$\,mag, 
UKIDSS DR9 $Y = 18.039$\,mag, $J = 17.898$\,mag, $H = 17.580$\,mag, 
and $K = 17.612$\,mag \citep{lawrence07}. 
The fitting parameters encompass a distance scaling factor, a measure 
for the interstellar extinction 
(parameterized here according to the description of 
\citealt{fitzpatrick99}), and the surface ratio of the two stars, 
which is needed to weight the fluxes in the composite SED. 
While atmospheric parameters for the sdB {\sc Tlusty} model are kept fixed 
and are taken from spectroscopy, the temperature 
and surface 
gravity of the cool companion are fitted with models extracted from 
the {\sc Phoenix} grid of \cite{husser13} as the {\sc BlueRed} grid does not
cover the SED. 
This way we also test whether the results for the K dwarf are model-dependent. 
To enhance the sensitivity of the photometric analysis to $\log{g}_2$, 
we express the surface ratio, which is well-constrained because the 
SED covers data points dominated by the sdB (ultraviolet), as well as by 
the K-star (infrared), as function of gravity and mass, 
i.e. $(R_2/R_1)^2 = M_2 g_1 /(M_1 g_2)$ 
with $M_1$ set to the canonical 
mass of $0.48\,M_\sun$ and $\log{g}_1$ to its spectroscopic value. 
Given the K3V type of the companion, its mass is limited to 
the small range $0.7$--$0.9\,M_\sun$ effectively making the surface 
ratio a probe for $g_2$. 
The results of the fit are listed in Table~\ref{tbl:one} and 
visualized in Fig.~\ref{fig:2} and show a perfect agreement with 
spectroscopy.
The given uncertainties are statistical $1\sigma$-confidence 
intervals based on the $\chi^2$ statistics of the residuals 
shown in Fig.~\ref{fig:2}.
Note that the offset with respect to the spectroscopic 
distance is mainly caused by discrepancies in the absolute 
flux calibration between the SDSS spectrum and photometry, 
which are of the order of $0.1$\,mag. 

\section{Kinematics\label{kinematics}}

\begin{figure*}
\centering
\includegraphics[width=0.8\textwidth]{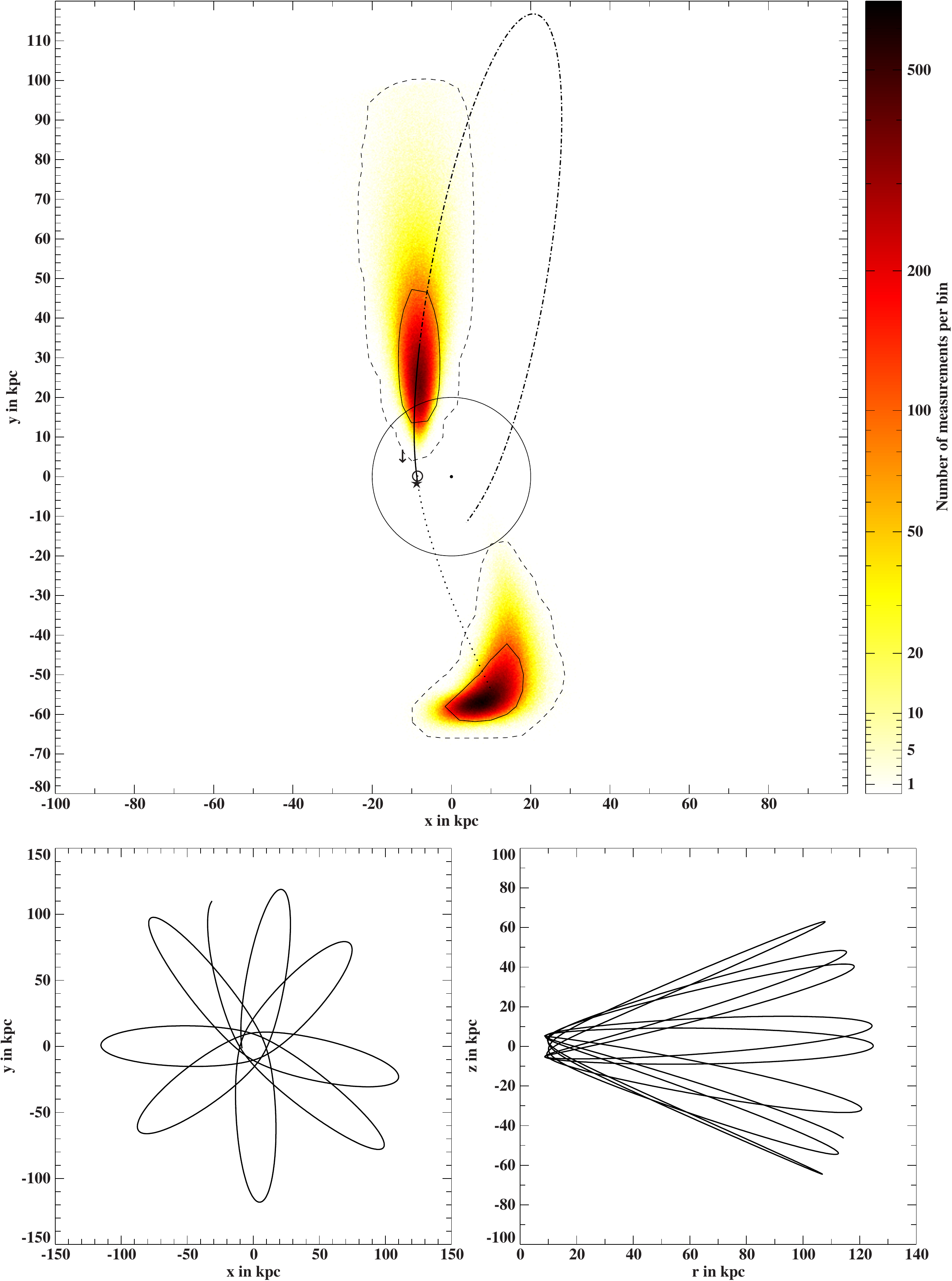}
\caption{
{\sl Top panel:} disk passages (binned and color coded) of J1211 from our Monte Carlo
simulation using Model-III of \cite{irrgang13} with 1$\sigma$ (solid) and
3$\sigma$ (dashed) contours. 
The solar symbol marks the position of the Sun, 
the black dot represents the position of the Galactic Center and the star
marks the current position of J1211. 
The projection of one orbit onto the $x$-$y$-plane is shown with the
dashed-dotted line for $z<0$, with a full line from $z=0$ to the current position, and 
with a dotted line from the current position of 
J1211 to the next passage of the 
$x$-$y$-plane. 
{\sl Bottom panels:} estimated orbits of J1211 in the $x$-$y$-plane (left) and
$r$-$z$-plane (right) in the most massive Galactic potential over the past 10 Gyr.
\label{fig:3} }
\end{figure*}

To obtain the dynamical properties of J1211, three different Milky Way mass
models (see models I, II, and III by \citealt{irrgang13}) were used to trace
back the orbit to the Galactic disk. 
Model-I is a revised model of \citet{allen91} with a halo mass of
M$_{\rm R<200\ kpc} = 1.9\times10^{12} M_\odot$ within a radius of 200 kpc.
For Model-II, the halo mass distribution is replaced by a truncated, flat
rotation curve model according to \citet{wilkinson99} and \citet{sakamoto03} 
with M$_{\rm R<200\ kpc}=1.2\times10^{12} M_\odot$. 
The dark matter halo in Model-III has 
M$_{\rm R<200\ kpc}=3.0\times10^{12} M_\odot$ and is based on the density
profile by \citet{navarro97} derived from cosmological simulations.
All mass models are the sum of a central spherical bulge component, an
axisymmetric disk component and a massive spherical dark matter halo. 
The model parameters were recalibrated using new and improved observational
constraints by \citet{irrgang13}.
The intersection area of the trajectories of J1211 with the Galactic plane
and the median $v_{\rm grf}$ at the present location were
determined by varying the position and velocity components within their
respective errors by applying a Monte Carlo procedure with a depth of $10^6$
for each mass model. 
The bound probability is defined as the number of orbits not exceeding the
local escape velocity, i.e. the number of orbits with a negative sum of
their potential and kinetic energy, with respect to the number of all 
calculated orbits.
 
\begin{figure}
\centering
\includegraphics[width=\linewidth]{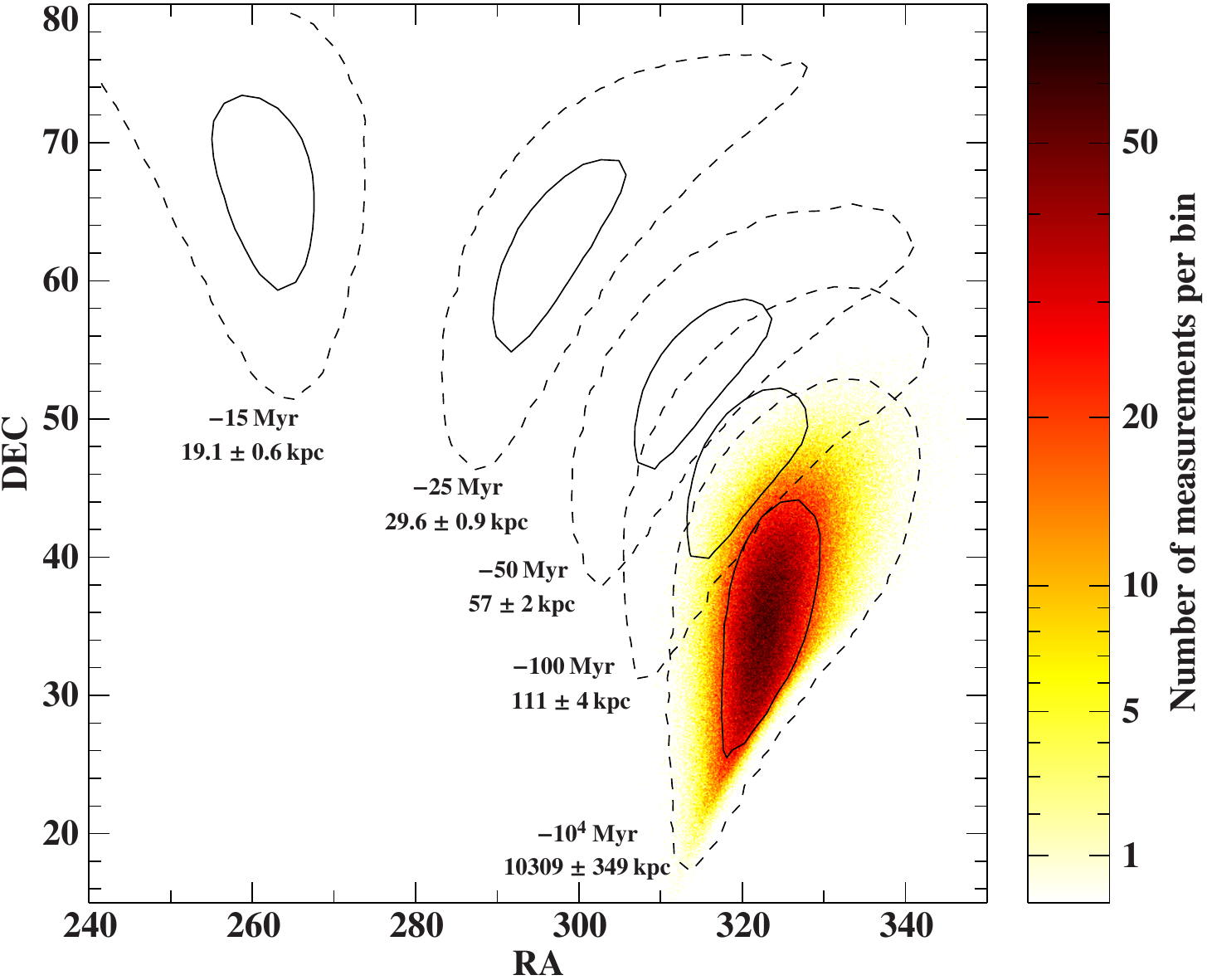}
\caption{ 
The sky locations and Galactic Center 
distances of J1211 in the past.
These are calculated from the unbound orbits in the least massive Galactic mass model.
\label{fig:4} }
\end{figure}

Using the combined spectroscopic and photometric distance of $d=5.5\pm0.5$ kpc 
we found J1211 at a Galactocentric distance of $10.2\pm0.3$ kpc 
with $v_{\rm grf}=571.3\pm76.4$ km\,s$^{-1}$.
To reconstruct the orbit we took proper motions from \citet{tillich11}. 
For the most massive mass model (Model-III) 99.7\% of the orbits are bound to
the Galaxy; for the least massive mass model (Model-II) only 40.3\% of the
orbits are bound to the Galaxy. 
Model-I gave a probability of bound trajectories of 62.4\%. 
In Fig.~\ref{fig:3} we show that an origin of the system 
from the Galactic Center can be excluded consistently for all three models.
The current location of J1211 is very close to the perigalacticon point
of its orbit in the most massive model.
We list the parameters of this orbit in Table~\ref{tbl:one}.
Fig.~\ref{fig:4} shows the past locations of J1211 in the equatorial
coordinate system assuming an unbound orbit.
 
{\sl Gaia} \citep{perryman01} will provide parallax measurements for J1211 
($G=17.96$, $V-I=-0.013$ mag), with a standard error 
of $\sigma_\pi=132\,\mu{as}$. 
With its $\pi=182^{+18}_{-15}\,\mu{as}$ parallax (from $d=5.5\pm0.5\,kpc$) J1211
is too faint and too far for {\sl Gaia} to improve on our distance measurement. 
However, with the anticipated $80\,\mu{as}$ end-of-mission proper motion 
standard error {\sl Gaia} will provide about 20 times lower error for the proper
motion and improve our kinematic parameters and bound probabilities.

\section{Conclusions \label{discussion}}

Our analysis has lowered the distance and tangential velocity 
of J1211 by 2$\sigma$ compared to \cite{tillich11}. 
Thanks to the new radial velocity measurements and improved distance 
we also achieved a 
lower uncertainty on the Galactic rest-frame velocity.
The unexpected discovery of a cool companion in a wide orbit around J1211
has severe consequences when trying to explain the extreme kinematics of
this object.
Along with the SN ejection scenario, which predicts the ejection of a
single hot subdwarf star, essentially all other acceleration mechanisms
discussed for HVS and runaway stars can be immediately excluded.
A wide binary system like J1211 is too fragile to survive any kind of
close encounter with another object that might lead to such an acceleration.
A close encounter with a supermassive or intermediate-mass black hole 
would rip the binary apart.
The same would happen during a dynamical interaction in a dense stellar
population or could be caused by the kick of a core-collapse SN.
Only very close binaries 
might survive such interactions.
A globular cluster tidal tail origin is not likely either, as it would
require a high-velocity cluster, and due to the low binary fraction in
globular clusters we would expect more single HVS stars to be
on similar orbits than binaries.

Should J1211 be unbound to the Galaxy, the origin of the system is an
interesting question.
Half of the known HVS stars concentrate 
in the direction of the constellation Leo \citep{brown14}. 
J1211 is in this direction, but its velocity vector is perpendicular to
those of the known early-type HVS stars, suggesting a different origin.
Although we could not associate J1211 with any of the known stellar
streams, it may be the first evolved HVS system discovered from the 
tidal debris of a disrupted dwarf satellite galaxy.
 
If J1211 has a bound orbit, it must have formed 
with the Galaxy or accreted later.
Due to the extreme kinematics of the system, a bound orbit requires a 
massive dark matter halo, therefore J1211 is particularly well suited 
to probe the Galactic potential and constrain the mass of the dark 
matter halo as described in \cite{przybilla10}.

We conclude that the kinematic properties of J1211 are most likely
primordial and that the binary must be an extreme halo object, which was
either born in the halo population or accreted later from the debris
of a destroyed satellite galaxy.
The low metallicity of the K-star companion is consistent with both
scenarios.
The progenitor of the sdB was most likely a main-sequence star more massive
than the halo
turnoff mass of about $0.8\,M_{\rm \odot}$.
Such old population II stars orbiting through the outermost
parts of our Galaxy with extreme kinematics are very rare
(\citealt{tillich10}; \citealt{scholz15}).

\begin{acknowledgements}
P.N. and E.Z. were supported by the Deutsche Forschungsgemeinschaft (DFG)
through grants HE1356/49-2 and HE1356/45-2, respectively. T.K. 
acknowledges support by the Netherlands Research School for Astronomy
(NOVA).
This work is based on observations at the La Silla-Paranal Observatory of the European    
Southern Observatory for programme number 093.D-0127(A). 
This work is based on observations obtained at the W.M. Keck Observatory, which is
operated as a scientific partnership among the California Institute of
Technology, the University of California, and the National Aeronautics and
Space Administration. 
The Observatory was made possible by the generous financial support of the
W.M. Keck Foundation. 
This research has used the services of {\sc astroserver.org}.
\end{acknowledgements}

\end{document}